\documentclass{aa} % for a paper on 1 column  

\usepackage{graphicx}
\usepackage{color}

\begin{document}

%\title{Origin of SMBHs: \\ The discriminant between quasi-stars and supermassive stars}
%\title{Black hole formation in the early Universe: \\ Impact of the accretion rate on the outcome of collapse}
%\titlerunning{Impact of the accretion rate on the outcome of collapse}

%\title{Progenitors of supermassive black holes in rapidly accreting primordial halos}
\title{Massive black hole factories:  Supermassive and quasi-star formation in primordial halos}
\titlerunning{Massive black hole factories} 

\author
  {Dominik R.\,G. Schleicher
  \inst{1}
  \and
  Francesco Palla
  \inst{2}
  \and 
  Andrea Ferrara
  \inst{3, 4}
  \and
  Daniele Galli
  \inst{2}
  \and 
  Muhammad Latif
  \inst{1}
  }

\institute{Institut f\"ur Astrophysik, Georg-August-Universit\"at G\"ottingen, Friedrich-Hund-Platz 1, D-37077 G\"ottingen, Germany \\
\email{dschleic@astro.physik.uni-goettingen.de, mlatif@astro.physik.uni-goettingen.de}
\and
INAF-Osservatorio Astrofisico di Arcetri, Largo Enrico Fermi 5, I - 50125 Firenze, Italy\\
\email{palla@arcetri.astro.it, galli@arcetri.astro.it}
\and
Scuola Normale Superiore, Piazza dei Cavalieri 7, 56126 Pisa, Italy\\
\email{andrea.ferrara@sns.it}
\and
Kavli IPMU (WPI), the University of Tokyo, 5-1-5 Kashiwanoha, Kashiwa, 277-8583, Japan
}

\date{\today}

%\maketitle

\abstract
{Supermassive stars and quasi-stars (massive stars with a central black hole) are both considered as potential progenitors for the formation of supermassive black holes. They are expected to form from rapidly accreting protostars in massive primordial halos.}
{We explore how long rapidly accreting protostars  remain on the Hayashi track, implying large protostellar radii and weak accretion luminosity feedback. We  assess the potential role of energy production in the nuclear core, and determine what regulates the evolution of such protostars into quasi-stars or supermassive stars.}
{We follow the contraction of characteristic mass scales in rapidly accreting protostars, and infer the timescales for them to reach nuclear densities. We compare the characteristic timescales for nuclear burning with those for 
which the extended protostellar envelope can be maintained.}
{We find that the extended envelope can be maintained up to protostellar masses of $3.6\times10^8\,\dot{m}^3$~M$_\odot$, where $\dot{m}$ denotes the accretion rate in solar masses per year. We expect the nuclear core to exhaust its hydrogen
content in $7\times10^6$~yr. If accretion rates $\dot{m}\gg 0.14$ can still be maintained at this point, a black hole may form within the accreting envelope, leading to a quasi-star. Alternatively, the accreting object will gravitationally contract to become a main-sequence supermassive star. }
{Due to the limited gas reservoir in typical $10^7$~M$_\odot$ dark matter halos, the accretion rate onto the central object may drop at late times, implying the formation of supermassive stars as the typical outcome of direct collapse. However, if high accretion rates are maintained, a quasi-star with an interior black hole may form.}

\maketitle

\section{Introduction}

Supermassive black holes (SMBHs) with more than $10^9$~M$_\odot$ have been observed at $z>6$ \citep{Fan01, Fan03, Fan04, Fan06}, and recently even at $z=7.085$ \citep{Mortlock11}. While theoretical considerations indicate a variety of potential pathways, including direct collapse to a black hole, collapse to a supermassive star, collapse of stellar clusters or clusters of black holes \citep{Rees84, Haiman06, Volonteri12}, substantial efforts have been made to assess the validity of such scenarios. As the most straightforward possibility, one could imagine these SMBHs to originate from remnants of the first stars, which were proposed to be very massive with $100-1000$~M$_\odot$ \citep{Abel02, Bromm04}. Stellar evolution calculations by \citet{Heger02} indeed show that black holes may form for stellar masses between $10-30$~M$_\odot$, or $100-300$~M$_\odot$, thus yielding the seeds for further growth to SMBHs.

New calculations following the evolution of primordial star formation beyond the first peak however show that fragmentation can occur efficiently \citep{Stacy10, Turk10, Clark11, Greif11, Smith11, Stacy12}. While the effect of the accretion luminosity on the protostellar core seems to be minor \citep{Smith11, Smith12}, UV feedback during the main sequence phase appears to set
the upper limit to the stellar mass to $50-100$~M$_\odot$ \citep{Hosokawa11, Hosokawa12b, Susa13}. Even if very massive stars could form, the resulting HII region is likely to inhibit accretion for at least $10^8$~yrs \citep{Milos09a, Milos09b}. An additional problem has been identified by \citet{Whalen12}, who showed that black holes resulting from such stars would be born with strong dynamical kicks, implying that they should readily be expelled from the halo.

On the other hand, dense stellar clusters may form  at high redshift in the presence of trace amounts of dust \citep{Schneider03, Omukai05, Clark08, Omukai08, Dopcke11, Schneider11, Dopcke12, Klessen12, Omukai12, Schneider12}. The relativistic instability of such clusters has been explored in a semi-analytic framework by \citet{Devecchi09} and \citet{Devecchi10, Devecchi12}, finding characteristic black hole masses of up to $\sim 3000$~M$_\odot$. The latter is already substantially higher than the masses of the first stars,  though it may still be difficult to obtain masses as high as $10^9$~M$_\odot$ at $z=7$. In the presence of a significant black hole spin, which will be rapidly obtained in the presence of efficient accretion, one typically expects an additional growth by about four orders of magnitude, thus requiring seed masses of the order $10^5$~M$_\odot$ \citep{Shapiro05}.

The formation of such massive seeds has been considered in the context of the direct collapse model, where gas in a $10^7$--$10^8$~M$_\odot$ halo is expected to collapse without fragmentation into a single central object \citep{Koushiappas04, Begelman06, Lodato06, Spaans06, Volonteri08, Begelman09}. In order to avoid fragmentation, molecular hydrogen needs to be efficiently dissociated, requiring a strong ambient UV field \citep{Omukai01, Bromm03, Shang09, Schleicher10b, Latif11b, Borm13}. Such radiation backgrounds may have been provided by nearby starburst galaxies, which may indeed occur frequently enough to explain the observed abundance of SMBHs at $z\sim6$ \citep{Dijkstra08, Agarwal12}. 

Using numerical simulations, \citet{Wise08a} have modeled the gravitational collapse of massive primordial halos cooling via atomic hydrogen, reporting an isothermal density profile and angular momentum transport via bar-like instabilities during the formation of the first peak. \citet{Regan09} presented the first study following the evolution beyond the first peak, and modeling the formation of self-gravitating disks on parsec-scales. While these studies employed a typical resolution of $16$ cells per Jeans length, it was recently demonstrated that turbulent structures can only be resolved with a resolution of at least $32$ cells per Jeans length, preferably more \citep{Sur10, Federrath11, Turk12}. In a set of high-resolution simulations following the formation of the first peak, \citet{Latif13a, Latif13b} found extended turbulent structures in the center of massive primordial halos, but no signs of simple disks or bar-like instabilities. Disks do however form at later stages of the evolution, with masses 
of $\sim1000$~M$_\odot$ on scales of $30$~AU, and characteristic accretion rates of $\sim1$~M$_\odot$~yr$^{-1}$ \citep{LatifBH}. As a result, very massive central objects may indeed form in rather short cosmic times. The dynamics at early times have also been explored by \citet{Choi13}, finding a somewhat different result including the formation of toroidal structures on scales of several parsec. It is however not fully clear to which extent the latter is a result of the simplified initial conditions employed in their calculation. In order to provide more quantitative predictions regarding the conditions where black holes may form, \citet{Prieto13} explored the correlations of the baryon spin with the spin of the dark matter, showing that the resulting correlation is however weak, and that the baryon properties cannot be naively extrapolated from the dark matter.

While the accumulation of high masses seems feasible from a hydrodynamical point of view, the resulting object is still considerably more uncertain. As a first step, we expect the formation of a massive protostars, as the gas becomes optically thick at high densities \citep{Omukai01prot, Omukai03, Hosokawa12}. It is however unclear how these objects are going to evolve, and whether they will form a supermassive star or a quasi-star. While a supermassive star denotes a conventional star with very high masses of $10^3-10^6$~M$_\odot$ \citep{Shapiro86}, a quasi-star refers to an object with similar mass, but where the central core has collapsed into a low-mass black hole \citep{Begelman06, Begelman10}. The latter then accretes mass from the stellar envelope, while the quasi-star as a whole may accrete at a rate larger than the Eddington rate of the black hole. A central question however concerns the conditions under which we expect the formation of a quasi-star as opposed to a supermassive star, and which of 
these objects should be expected as the generic outcome in the context of the direct collapse model. Also the properties of these objects are of high interest, as the amount of stellar feedback may influence the final accretion rates.

In order to explore that question, \citet{Hosokawa12} recently followed the  evolution of rapidly accreting protostars, showing that they expand as cool supergiants, thus inhibiting the feedback from accretion luminosity and potentially allowing accretion to proceed for very long times. Hosokawa et al. also found that nuclear burning starts at stellar masses of about $\sim50$~M$_\odot$, and is maintained until the end of their calculation at $10^3$~M$_\odot$. Therefore, their model indicates no transition towards a quasi-star at least during the evolutionary phase considered. It is however important to assess how long this phase of efficient accretion can be maintained, and under which conditions a transition to a supermassive main-sequence star can be expected, which can potentially influence the accretion flow via radiative feedback \citep{Omukai02, Johnson11, Johnson12}. While the majority of such supermassive stars directly collapses into a black hole \citep{Fryer11}, a small mass window exists around 
$55000$~M$_\odot$ where violent supernova explosions of up to $10^{55}$~erg may occur \citep{Johnson13}.

Following a different approach, \citet{Begelman06} and \citet{Begelman10} have proposed the formation of a black hole in the interior of rapidly accreting objects, leading to quasi-stars as the progenitors of SMBHs.  \citet{Begelman06} argued that for quasi-stars with $10^4$--$10^5$~M$_\odot$,  the typical accretion timescale is considerably shorter than the timescale for nuclear burning, and as a result, the latter may not be able to stop the collapse of the central core. A potential advantage of such a configuration is that the accretion onto the central object is not limited by the Eddington accretion rate of the black hole, but by the Eddington accretion rate of the more massive quasi-star \citep{Begelman08}.   \citet{Begelman10} considers objects with more than $10^6$~M$_\odot$, which first evolve towards the main sequence as a supermassive star, but then form a black hole after an extended phase of nuclear burning. \citet{Ball11} followed the evolution of black holes in a quasi-star employing the Cambridge STARS stellar evolution package \citep{Eggleton71, Pols95}, finding that the black hole may efficiently accrete $10\%$ of the stellar mass before hydrostatic equilibrium breaks down. They also report that the results are sensitive to the boundary condition in the interior. The contraction of an embedded isothermal core in a stellar envelope was further explored by \citet{Ball12} in the framework of the Sch\"onberg-Chandrasekhar limit.

While the models for the quasi-stars typically employed a Thompson-scattering opacity, \citet{Hosokawa12} reported that  H$^-$ rather than Thomson scattering dominates the opacity in the protostellar atmosphere. They find luminosities  close to the Eddington luminosity,\begin{equation}
L_{\rm Edd}=\frac{4\pi G M m_p c}{\sigma_T}=3.8\times10^4L_\odot \left( \frac{M}{M_\odot} \right),\label{LEddstar}
\end{equation}
where $G$ is the gravitational constant, $m_p$ the proton mass, $c$ the speed of light, $\sigma_T$ the Thomson scattering cross section and $M$ the mass of the star. The temperatures in the atmosphere are rather cool, $\sim5000$~K, as the protostars are on the Hayashi track. As a result, the characteristic radii evolve as\begin{equation}
R_{ini}=2.6\times10^2R_\odot\left( \frac{M}{M_\odot} \right)^{1/2}.\label{Rint}
\end{equation}
The protostars are thus considerably more extended and show an increasing stellar radius as a function of mass, while the models of \citet{Begelman06} and \citet{Begelman10} indicate a constant radius as a function of stellar mass. The difference is crucial, as the stellar radius regulates the temperature on the surface, and thus the strength of protostellar feedback.
In fact, the behavior reported by \citet{Hosokawa12} is well-known in cases where the accretion timescale,
\begin{equation}
t_{acc}=\frac{M}{\dot{M}},
\end{equation}
is much smaller than the Kelvin-Helmholtz timescale,
\begin{equation}
t_{KH}=\frac{GM^2}{R L}.
\end{equation}
However, \citet{Hosokawa12} report that this behavior extends into the regime where $t_{KH}<t_{acc}$, i.e. where Kelvin-Helmholtz contraction is faster than mass growth via accretion.  The physical mechanism which allows this phase to continue will be discussed in detail in section~\ref{KH}. Taking the maximum luminosity of the star
\begin{equation}
L_{max}\sim0.6\,L_\odot \left( \frac{M}{M_\odot} \right)^{11/2}\left( \frac{R}{R_\odot} \right)^{-1/2},
\end{equation}
which assumes Kramer's opacity $\kappa\propto \rho T^{-3/5}$ \citep{Hayashi62} and an initial radius as given in Eq.~(\ref{Rint}), they show that both timescales are equal at protostellar masses of
\begin{equation}
M_{eq}=14.9\,M_\odot\left( \frac{\dot{m}}{0.01}  \right)^{0.26},
\end{equation}
where we parametrized the accretion rate as $\dot{M}\equiv \dot{m}$~M$_\odot$~yr$^{-1}$. On the other hand, the accreting star remains on the Hayashi track up to stellar masses of $1000$~M$_\odot$, the highest mass reached in the calculation of \citet{Hosokawa12}. 
The extended envelope locked at $T_{\rm eff}\sim 5000$~K allows efficient accretion with only moderate feedback for a longer period. 

In this paper, we aim to assess how long this efficient accretion phase can be maintained beyond $M\sim 1000$~M$_\odot$ without strong feedback from the protostar. In addition, we will discuss the potential impact of nuclear burning for the evolution of the accreting objects, and the conditions under which a quasi-star as opposed to a supermassive star may form. For this purpose, we explore the interplay of mass accretion and Kelvin-Helmholtz contraction in section~\ref{KH}. We calculate the impact of nuclear burning in section~\ref{nuclear}. A final discussion of our results is provided in section~\ref{discussion}. The impact of additional processes such as deuterium shell burning and hydrogen burning via the pp-chain is assessed in the appendix.

\begin{figure}[t]
\begin{center}
\includegraphics[scale=0.52]{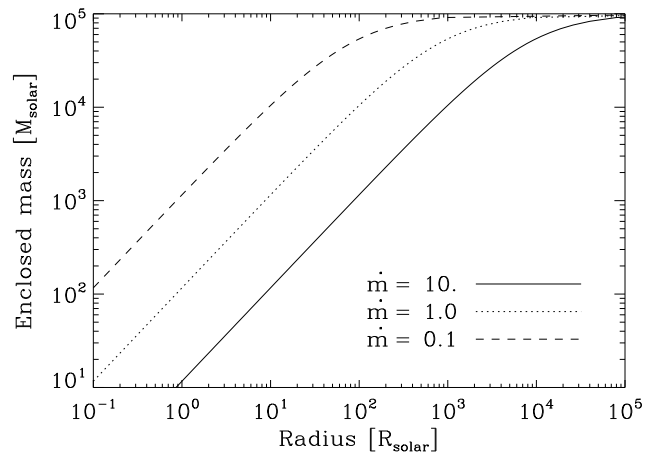}
\caption{The enclosed mass as a function of radius in the protostar calculated from Eq.~(\ref{evR}), assuming a total stellar mass of $10^5$~M$_\odot$ and mass accretion rates $\dot{m}=0.1,1.,10$. The current time $t$ is calculated as $t=(M/M_\odot)/\dot{m}$. For a given mass shell, larger accretion rates imply larger radii, as the star had less time to contract. }
\label{radius}
\end{center}
\end{figure}

\section{The interplay of mass accretion and Kelvin-Helmholtz contraction}\label{KH}
In the following, we will sketch why the protostar may maintain large envelopes even if the Kelvin-Helmholtz timescale becomes shorter than the timescale for accretion. For this purpose, we consider mass shells of enclosed mass $M$, located at radii $R(M,t)$. A mass shell $M$ forms at the time $M/\dot{M}$, with an initial radius as given by Eq.~(\ref{Rint}). The radii of these mass shells evolve on their Kelvin-Helmholtz timescale, given as
\begin{equation}
t_{KH}(M,R)=\frac{GM^2}{R(M,t) L_{\rm Edd}}.
\label{tkh}
%\sim2.7\times10^7\mathrm{s}\frac{M/M_\odot}{R/(1000 R_\odot)}.\label{KH}
\end{equation}
As a result, we have\begin{equation}
\frac{dR}{dt}=-\frac{R(M,t)}{t_{KH}(M,R)}.
%\sim-2.6\times10^6\frac{cm}{s} \frac{(R/(1000 R_\odot))^2}{M/M_\odot}.\label{dRdt}
\end{equation}
The equation is integrated between the initial radius $R_{ini}$ when the mass shell forms, given in (\ref{Rint}), corresponding to the time $t_{ini}(M)=M/\dot{M}$, to the current radius $R$ at time $t$. The integration yields
\begin{equation}
\frac{1}{R}=\frac{1}{R_{ini}}+\frac{4\pi m_p c}{\sigma_T M}(t-t_{ini}(M)).\label{shell}
\end{equation}
Inserting the initial radius as given in Eq.~(\ref{Rint}) and expressing the result in astrophysical units, we obtain\begin{equation}
\frac{1000 R_\odot}{R}=\frac{1000 R_\odot}{2.6\times10^2(M/M_\odot)^{1/2}}+\frac{1.04\,\mathrm{yr}^{-1}\left(t-t_{ini}(M)\right)}{M/M_\odot}.\label{evR}
\end{equation}
The first term on the right-hand side thus dominates right after the formation of a mass shell and determines its initial radius, while the second term describes its evolution due to Kelvin-Helmholtz contraction. It is remarkable that at late times, the second term dominates and the evolution of the radii appears to be independent of their initial position. This can be understood, as our expression for the Kelvin-Helmholtz timescale, Eq.~(\ref{tkh}), scales with the inverse radius of the mass shell. As a result, the evolution slows down during the contraction, implying that the Kelvin-Helmholtz timescale may increase significantly in the interior. An example for the resulting structure in a $10^5$~M$_\odot$ star is given in Fig.~\ref{radius}, including both the regime where the first and the second term of the equation dominate. 

We now aim to estimate when a given mass shell reaches the densities of nuclear burning, which we take as $\rho_{nuc}\sim1$~g~cm$^{-3}$ \citep{Hosokawa12}. We obtain a radius of nuclear burning, which is given as
\begin{equation}
R_{nuc}=\left( \frac{3M}{4\pi\rho_{\rm nuc}} \right)^{1/3}\sim\ 1.2R_\odot\left(\frac{M}{M_\odot}\right)^{1/3}.\label{Rnuc}
%\sim1.2R_\odot(M/M_\odot)^{1/3}.\label{Rnuc}
\end{equation}
A comparison with Eq.~(\ref{Rint}) shows that this radius always remains smaller than the radius of the star by at least a factor of $1000$, and in fact increases more gradually with stellar mass. A given mass shell $M$ will only reach nuclear densities on timescales much longer than the initial Kelvin-Helmholtz timescale $t_{KH}$, which increases as $R^{-1}$.  We can thus determine the time when nuclear densities are reached by equating (\ref{Rnuc}) with the second term in (\ref{evR}). As a result, we obtain\begin{equation}
\Delta t=t-t_{ini}(M)=710\ \mathrm{yrs}\left( \frac{M}{M_\odot} \right)^{2/3}.\label{Deltat}
\end{equation}
During the time interval $\Delta t$, the star will accrete an additional mass $\Delta M=\dot{M} \Delta t$. Considering a given mass shell, we can  calculate the ratio \begin{equation}
\frac{\Delta M}{M}=\frac{\dot{M}\Delta t }{M}\sim710\,\dot{m}\left( \frac{M}{M_\odot} \right)^{-1/3},
\end{equation}
which describes how much additional mass is accreted before the shell reaches nuclear densities. The ratio drops below $1$ only for protostellar masses of \begin{equation}
M\geq3.6\times10^8\,\dot{m}^3M_\odot.\label{Mmax}
\end{equation}
While for typical accretion rates of $\dot{m}\sim10^{-3}$, this happens already at mass scales of order unity, this transition occurs only at substantially larger masses above $1000$~M$_\odot$ for $\dot{m}>10^{-2}$. While a given shell evolves towards nuclear densities, the amount of additional mass that is accreted is thus substantially higher than the  mass in that shell. From Eq.~(\ref{Deltat}), we can further derive the mass in the nuclear core in the limit that $t\gg t_{ini}(M)$. We obtain\begin{equation}
M_{nuc}=\left( \frac{t}{710\ \mathrm{yr}}  \right)^{3/2}M_\odot=t_{710}^{3/2}~M_\odot,
\end{equation}
where we defined $t_{710}=t/(710$~yr$)$. As we neglected the term $M/\dot{M}$ in Eq.~(\ref{Deltat}), we note that the derivation here implicitly assumes a sufficient mass supply to the protostar in order to feed the core. Clearly, this approximation will break down at the mass scale derived in (\ref{Mmax}). At that point, both the protostellar mass as well as the mass in the nuclear core may become approximately constant, and we expect a transition towards supermassive main sequence stars.

From this model, we already obtain a set of relevant conclusions concerning the evolution of the protostar. In particular, if the accretion rate is high, a substantial amount of additional matter is accreted before a given shell reaches nuclear densities. As a result, the evolution of the protostar is dominated not by the interior shells, but by the additional matter which is accreted during the contraction. This behavior will only change when the timescale of accretion becomes comparable to the timescale on which the outer mass shell is able to reach nuclear densities. As we show here, the latter is significantly larger than the Kelvin-Helmholtz timescale of the star, and scales with the $R_{nuc}^{-1}$. The transition occurs when a critical mass scale of $3.6\times10^8\dot{m}^3$~M$_\odot$ is reached, implying a transition towards a supermassive main sequence star. We note that the transition may even occur earlier if $\dot{m}$ decreases with time.

\section{The nuclear evolution}\label{nuclear}
In this section, we will assess the potential role of nuclear burning. For this purpose, we show that the transition to the CNO-cycle rapidly occurs and regulates the production of helium in the nuclear core. We will then calculate under which conditions the fuel in the nuclear core will be exhausted while the protostar maintains its extended envelope. The latter implies the potential formation of a black hole in the interior, thus a transition into a quasi-star.

\begin{figure}[t]
\begin{center}
\includegraphics[scale=0.52]{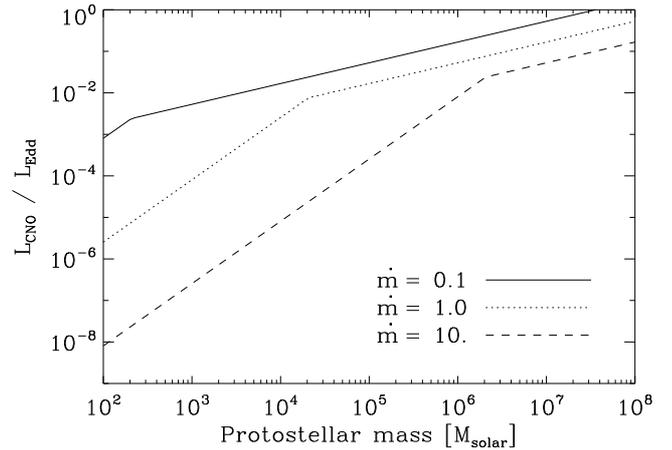}
\caption{The ratio of nuclear luminosity provided via the CNO-cycle over the Eddington luminosity of the star (\ref{EqLEdd}) as a function of protostellar mass for different mass accretion rates. The CNO luminosity is calculated here as the minimum of the estimate in Eq.~(\ref{EqCNO} and the Eddington luminosity of the nuclear core (\ref{EqCore}). At high protostellar masses, one can clearly recognize the transition to the regime where nuclear burning is limited by the Eddington luminosity of the core. Larger accretion times imply that less time was available for contraction, thus reduced nuclear luminosities.  }
\label{figCNO}
\end{center}
\end{figure}

\subsection{Importance of the CNO-cycle}\label{CNO}
While the pp-cycle will dominate in the very beginning (see appendix), we show here that the transition to the CNO-cycle rapidly occurs as a result of helium burning via the triple-$\alpha$ process. Following \citet{Padmanabhan00}, the energy production rate via triple-$\alpha$ is given as\begin{equation}
\epsilon_{3\alpha}=\frac{5.1\times10^8\rho^2Y^3}{T_9^3}e^{-4.4027/T_9}\mathrm{\ erg\ g}^{-1}\mathrm{\ s}^{-1},
\end{equation}
where $Y\sim0.25$ denotes the mass fraction of helium and $T_9$ denotes the gas temperature in units of $10^9$~K. It is straightforward to check that the energy production rate  changes by several orders of magnitude for temperature changes of the order $10\%$. As shown by \citet{Hosokawa12}, the highest temperatures with $T_9\sim0.15$ are expected in the center of the nuclear core, which has an enhanced nuclear density of $10$~g~cm$^{-3}$. With a volume filling factor $\epsilon_{\rm fill}\sim0.01-0.1$, the contribution of that region to helium burning is significant due to the steep temperature dependence. For the following estimates, we will assume that efficient mixing occurs throughout the nuclear core, implying that the heavy elements produced in the central region will be available throughout the core. On the other hand, if mixing is inefficient, nuclear burning via the CNO cycle will dominate even more in the central region due to its higher metallicity. 

Noting that the  energy released by a single triple-$\alpha$ reaction corresponds to $1.166\times10^{-5}$~erg, the heavy element production rate per unit volume is given as
\begin{equation}
n_{3\alpha}=\frac{\epsilon_{3\alpha}}{1.166\times10^{-5}~\mathrm{erg}}.
%=4.2\times10^{-4}\mathrm{\ g}^{-1}\mathrm{\ s}^{-1}.
\end{equation}
The CNO mass in the core thus evolves as
\begin{equation}
\dot{M}_{\mathrm{CNO}}=\dot{n}_{3\alpha}\epsilon_{\rm fill}M_{nuc}\times12 m_p,%=2.7\times10^{-19}\epsilon_{fill}\frac{M_\odot}{\mathrm{yr}}\left(\frac{t}{710\mathrm{yr}}  \right)^{3/2},
\end{equation}
and an integration yields\begin{equation}
M_{\mathrm{CNO}}=7.1\times10^{-17}\epsilon_{\rm fill}\,t_{710}^{5/2}\,M_\odot.
\end{equation}
Assuming efficient mixing, the resulting metallicity in the nuclear core is then\begin{equation}
Z=\frac{M_{\mathrm{CNO}}}{M_{nuc}}=6.3\times10^{-9}\,t_{710}\,\epsilon_{\rm fill}.\label{Zprod}
\end{equation}
We recall that the energy production rate in the CNO cycle is given as \citep{Padmanabhan00}\begin{equation}
\epsilon_{\mathrm{CNO}}=\frac{4.4\times10^{25}\rho X Z}{T_9^{2/3}}e^{-15.228/T_9^{1/3}}\mathrm{\ erg\ g}^{-1}\mathrm{\ s}^{-1}.
\end{equation}
Since the energy production rate depends sensitively on the temperature, we expect its contribution in the innermost core to be dominant. For the temperature $T_9=0.15$, it is straightforward to show that energy production via the CNO cycle becomes comparable to  the pp cycle for a metallicity of $Z_c=2\times10^{-12}$. A comparison with Eq.~(\ref{Zprod}) shows that the CNO cycle thus dominates after a short time of
\begin{equation}
t_{\mathrm{CNO}}=2.25\epsilon_{\rm fill,-1}^{-1}\mathrm{\ yr},
\end{equation}
where we introduced $\epsilon_{\rm fill}=0.1\epsilon_{\rm fill,-1}$. We note here that this timescale is probably not accurate, as our assumptions in section \ref{KH} (in particular concerning the Eddington luminosity) only become valid at later times. Nevertheless, this result implies that the CNO cycle can be expected to be relevant early on, and thus needs to be considered. As the production of elements becomes increasingly efficient at late times, we expect that the expression (\ref{Zprod}) will  yield a reasonable estimate in the period of interest.

With the above assumptions, the CNO luminosity due to nuclear burning is given as
\begin{equation}
L_{\mathrm{\mathrm{CNO}}}=\epsilon_{\mathrm{CNO},Z_c}\left(\frac{Z}{Z_c}\right)M_{nuc}\,\epsilon_{\rm fill},
\end{equation}
where $\epsilon_{\mathrm{CNO},Z_c}$ denotes the energy production rate via the CNO cycle evaluated at the critical metallicity $Z_c$. Inserting (\ref{Zprod}), we obtain\begin{equation}
L_{\mathrm{CNO}}=1.3\times10^5 \epsilon_{\rm fill}^2 \,t_{710}^{5/2}\,L_\odot.\label{EqCNO}
\end{equation}
The Eddington luminosity of the nuclear core is given as\begin{equation}
L_{{\mathrm{Edd,core}}}=3.8\times10^4\dot{m}\,t_{710}^{3/2}\,L_\odot.\label{EqCore}
\end{equation}
The ratio of these luminosities to the Eddington luminosity of the star is given in Fig.~(\ref{figCNO}). Even here, the luminosity produced by nuclear burning approaches the Eddington luminosity of the star only around stellar masses of $10^8$~M$_\odot$, implying no relevant impact on the stellar evolution during the earlier stages. A comparison of these expressions yields the timescale\begin{equation}
t_c=2.1\times10^4\epsilon_{\rm fill,-1}^{-2}\mathrm{yr}.
\end{equation}
For $\epsilon_{\rm fill,-1}\sim1$, as indicated by \citet{Hosokawa12} for $\dot{m}=0.1$, the CNO luminosity would exceed the Eddington luminosity of the core after $\sim2\times10^4$~yr, shortly after the end of their simulations. However, when the Eddington luminosity is reached, one would expect an expansion of the nuclear core, implying lower densities and an adiabatically decreasing temperature. Due to this 
thermostat, it is likely that the core will adjust to a state maintaining its Eddington luminosity.

We will now demonstrate that the nuclear core is not converted to helium before the critical timescale $t_c$ after which the core radiates at its Eddington luminosity. We consider the production of helium by the CNO process, given as\begin{equation}
\dot{N}_{\mathrm{He,CNO}}=\frac{L_{\mathrm{CNO}}}{4.3\times10^{-5}\mathrm{\ erg}},%=1.2\times10^{43}\epsilon_{fill}^2\left( \frac{t}{710\mathrm{\ yr}} \right)^{5/2}\frac{1}{s},
\end{equation}
and the helium mass production rate\begin{equation}
\dot{M}_{\mathrm{He,CNO}}=12m_p\times\dot{N}_{\mathrm{He,CNO}}.%=3.8\times10^{-6}\epsilon_{fill}^2\,t_{710}^{5/2}\, \frac{M_\odot}{\mathrm{yr}}.
\end{equation}
An integration yields\begin{equation}
M_{\mathrm{He,CNO}}=1.0\times10^{-3}\epsilon_{\rm fill}^2\,t_{710}^{7/2}\,M_\odot.
\end{equation}
A comparison with the mass of the core indicates that they become comparable at\begin{equation}
t_{\mathrm{He,CNO}}=2.25\times10^5\epsilon_{\rm fill,-1}^{-1}\mathrm{\ yr},
\end{equation}
implying that a pure helium core will only be formed once the nuclear luminosity is equal to the Eddington luminosity of the core.

\begin{figure}[t]
\begin{center}
\includegraphics[scale=0.52]{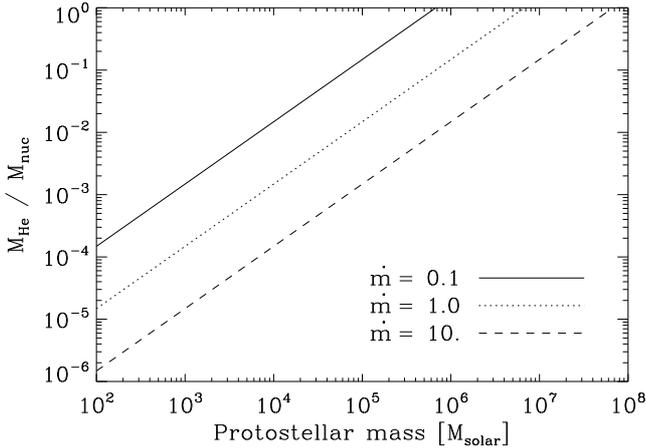}
\caption{The fraction of helium in the nuclear core, corresponding to the amount of exhausted fuel for nuclear burning, as a function of protostellar mass for different accretion rates. Higher accretion rates imply less time for protostellar contraction. As a result, nuclear densities are reached later, implying less time for nuclear burning.  }
\label{He}
\end{center}
\end{figure}

\subsection{Formation of a helium core}\label{heliumcore}

We will now calculate the helium production rate under the assumption that the formation of a helium core occurs 
once the nuclear luminosity equals the Eddington luminosity of the core. Thus,
\begin{equation}
\dot{N}_{\rm He}=\frac{L_{\mathrm{Edd,core}}}{4.3\times10^{-5}\mathrm{\ erg}}.%=3.5\times10^{42}\frac{1}{s}\left( \frac{t}{710\mathrm{\ yr}} \right)^{3/2}.
\end{equation}
The helium mass production rate is thus given as\begin{equation}
\dot{M}_{\rm He}=4m_p\times\dot{N}_{He}%=3.7\times10^{-7}\frac{M_\odot}{\mathrm{yr}}\left( \frac{t}{710\mathrm{\ yr}} \right)^{3/2},
\end{equation}
and an integration yields\begin{equation}
M_{\rm He}=1.05\times10^{-4}\,t_{710}^{5/2}\,M_\odot .
\end{equation}
Under this assumption, the nuclear core becomes a pure helium core after $t_f\simeq 6.8\times10^6$~yr.  We note that at this point, the mass of the nuclear core is given as $M_f\simeq 9.4\times10^5$~M$_\odot$. In order to illustrate how the latter translate into total stellar masses, the ratio of helium mass to total mass in the core is displayed in Fig.~\ref{He} for different accretion rates, showing that the hydrogen fuel is indeed exhausted for stellar masses of $3\times10^5-3\times10^7$~M$_\odot$ depending on the accretion rate.  After that point, the burning of heavier elements will occur, but only last for a short period. As the mass of the core is considerably larger than the Chandrasekhar mass scale and the Tolman-Oppenheimer-Volkoff mass scale, the core may collapse and form a black hole within the star, a so-called quasi-star, as proposed by \citet{Begelman06} and \citet{Begelman10}. The interior black hole may then continue accreting from the stellar envelope, while the accretion rate of the quasi-star will 
only be limited by the Eddington rate corresponding to the total mass of the configuration. The derivation of $t_f$ and $M_f$ however implicitly assumes that enough mass has been accreted in order to build up the core, requiring that
\begin{equation}
M_f\ll t_f \dot{M}.
\end{equation}
In order to form a quasi-star, we thus require an accretion rate of $\dot{M}\gg M_f/t_f$, i.e. \begin{equation}\dot{m}\gg 0.14.
\end{equation} 
For lower accretion rates, we expect that the extended envelope of the star cannot be maintained at late times, implying that the star will contract and form a supermassive main sequence star. During this period, both the stellar mass and the mass in the nuclear core will be approximately constant, implying a characteristic hydrogen burning timescale of $2\times10^6$~yr. For stellar masses above $300$~M$_\odot$, these are expected to collapse to massive black holes \citep{Fryer01, Fryer11}. 

\section{Discussion and outlook}
\label{discussion}

To investigate the interplay of accretion and Kelvin-Helmholtz-contraction of the protostar, we have presented an analytical model following the evolution of the mass shells from the extended envelope down to nuclear densities. Based on this model, we are able to explain the fundamental result by \citet{Hosokawa12} that the accreting protostar keeps its extended envelope beyond the adiabatic accretion phase. The reason is that the timescale on which a given mass shell may reach nuclear densities is much longer than the Kelvin-Helmholtz timescale of the star, as the protostellar radius is considerably larger than the radius of the nuclear core. Instead of the Kelvin-Helmholtz timescale evaluated at the protostellar radius, the relevant timescale is thus the Kelvin-Helmholtz timescale at the expected core radius after contraction, providing an appropriate estimate of the evolutionary timescale. As a consequence, rapidly accreting protostars are expected to keep their extended envelopes until stellar masses of $3.6\times10^8\,\dot{
m}^3$~M$_\odot$. At this point, the accretion timescale becomes longer than the Kelvin-Helmholtz timescale in the stellar interior, implying that the star will subsequently contract towards the main sequence. 
Then, we expect it to follow the typical evolution of a supermassive (proto)star \citep{Shapiro86}.

We further assessed the impact of nuclear burning for the evolution of these stars. While deuterium burning will never be relevant during the protostellar evolution, hydrogen burning via the pp-cycle could potentially affect the stellar evolution after $10^{15}$~years, which is however practically irrelevant. However, due to the efficient metal production via the triple-$\alpha$ process, a transition to the CNO cycle is expected early on. The energy production by the CNO cycle is considerably more efficient, implying that the core can radiate at its Eddington luminosity. As a result, a helium core forms after $7\times10^6$~yr. The expected mass of the core is then $10^6$~M$_\odot$, which provides an upper limit on the initial mass of the interior black hole.  Therefore, if the available gas reservoir is sufficient to maintain accretion rates $\dot{m}\gg 0.14$ until this point, the central core may collapse during the protostellar evolution phase, and the object becomes a quasi-star. Alternatively, the protostar will evolve towards the main sequence to become a supermassive star. Such an object is then expected to collapse into a supermassive black hole at the end of its lifetime \citep{Fryer11}.

Our results have been derived based on a comparison of the mass accumulated by accretion with the characteristic mass scale where we expect a pure helium core. This mass scale is based on the reasonable assumption that the luminosity of the core is given by its Eddington luminosity, in agreement with the numerical results by \citet{Hosokawa12}. Uncertainties are however present, as our calculation for the evolution of the core mass assumes that the luminosity of each mass shell is given by the Eddington luminosity. Following \citet{Hosokawa12}, the latter is a good but not precise approximation. Further corrections can be expected in particular when approaching nuclear densities, when feedback from nuclear burning becomes significant. Additional effects could be introduced as a result of rotation, which was not included in this model. We thus expect our results to provide an order of magnitude estimate for the critical accretion rate, which may be determined more accurately employing stellar evolution 
calculations, but also a more realistic model for the time-dependent accretion rate. We also note here that \citet{Hosokawa12} reported a slight dependence of their results on the employed boundary conditions for the protostar. While their main calculations adopted shock boundary conditions, they also explored the effect of lower-entropy accretion provided by photospheric boundary conditions. In that case, a slightly higher accretion rate of $0.3$~M$_\odot$~yr$^{-1}$ appears to be required to maintain the extended envelopes, but the overall evolution remains very similar.

For comparison, we note that recent simulations by \citet{LatifBH} reported accretion rates of $\sim1$~M$_\odot$~yr$^{-1}$ in halos with $\sim10^7$~M$_\odot$. Adopting a constant accretion rate, the transition to a supermassive main sequence star would occur at a stellar mass of $3.6\times10^8$~M$_\odot$. As the nuclear fuel is however exhausted earlier, a central black hole may form before this transition, giving rise to an evolution as sketched by \citet{Ball11}. However, due to the limited gas reservoir, and as here the halo mass is in fact comparable to the mass scale of the corresponding quasi-star, it seems likely that the accretion rate will substantially decrease at late times, implying that the transition towards the supermassive star should occur at an earlier stage which is then determined by the time evolution of the accretion rate, and that supermassive stars with $10^4-10^5$~M$_\odot$ might be the most generic outcome of the collapse. More massive objects appear to be possible at least in principle, if a larger gas reservoir is available in more massive dark matter halos.

In order to assess the potential influence of rotation, we estimated the amount of rotational support in the central $10^3$~M$_\odot$ clumps reported by \citet{LatifBH}, which varied between $5-20\%$ in different simulations. With densities of $\sim10^{-10}$~g~cm$^{-3}$, these are still orders of magnitudes below the characteristic densities within the protostar, such that we cannot yet draw strong conclusions regarding the final amount of rotational energy. However, we note that \citet{Stacy12} reported a significant amount of rotation in primordial protostars based on numerical simulations of \citet{Greif12}, and a quite similar case can be expected here. As recently shown by \citet{Reisswig13}, the latter may have a substantial impact on the collapse of supermassive stars, implying the potential formation of a black hole binary and a subsequent merger, accompanied with efficient emission of gravitational waves. The latter provides a potential pathway of probing black hole formation scenarios with LISA\
\footnote{LISA webpage: http://lisa.nasa.gov/}. In a narrow mass range around $\sim55000$~M$_\odot$, one may further expect the occurence of highly energetic supernovae with energies up to $10^{44}$~erg, which can be potentially detected with JWST\footnote{JWST webpage: http://www.jwst.nasa.gov/} \citep{Johnson13}.

Returning to the fate of supermassive stars, their evolution in the presence of UV feedback was assessed by \citet{Omukai02} and \citet{Johnson12} in the case of spherical symmetry. These authors find that UV 
feedback is unable to stop accretion for rates above $\sim 0.1$~M$_\odot$~yr$^{-1}$. As a result, the expected 
stellar mass is
\begin{equation}
M_{\rm UV}\sim10^3M_\odot\left( \frac{\dot{m}}{10^{-3}} \right)^{8/7}.
\end{equation}
For accretion rates of $\sim 0.1$~M$_\odot$~yr$^{-1}$, our results exceed this value, as the protostars remain on the Hayashi track up to a mass of $\sim 3\times10^5$~M$_\odot$, making feedback inefficient. Therefore, taking protostellar evolution into account favours the formation of more massive objects. In all cases, a critical question concerns the time evolution of the accretion rate, since the transition to supermassive stars is regulated by the accretion rate at late times. 
While this paper provides a first assessment for the case of constant accretion rates and spherical symmetry, the potential implications of time-dependent accretion rates need to be addressed in more detail in the future, along with the implications of rotation during protostellar evolution.

\begin{acknowledgements}
DRGS and ML thank for funding from the {\em Deutsche Forschungsgemeinschaft} (DFG) via the SFB 963/1 ``Astrophysical flow instabilities and turbulence'' (project A12). DRGS further acknowledges financial support via the {\em Schwerpunktprogramm} SPP 1573 ``Physics of the Interstellar Medium'' under grant SCHL 1964/1-1. DG and FP 
acknowledge the financial support of PRIN-INAF~2010 ``The Formation of Stars''.
\end{acknowledgements}

%\bibliography{astro}
%\bibliographystyle{aa}

\appendix

\section{Additional nuclear processes}

In section \ref{CNO}, we have shown that the transition to the CNO cycle rapidly occurs, implying that the latter will regulate the formation of a helium core as discussed in section \ref{heliumcore}. However, additional nuclear processes are expected to simultaneously occur, which we discuss here for completeness. In the two sub-sections below, we assess the role of deuterium shell burning as well as the impact of hydrogen burning via the pp-change during the stellar evolution, showing that these processes will only have a minor impact on the stellar evolution.

\subsection{Deuterium burning}
As reported by \citet{Hosokawa09}, deuterium shell burning may occur even before nuclear burning starts in the core. The luminosity from deuterium shell burning is given as\begin{equation}
L_D=1.5\times10^5L_\odot \left( \frac{\dot{m}}{0.1}   \right)\left( \frac{[\mathrm{D/H}]}{2.5\times10^{-5}} \right),
\end{equation}
where [D/H] denotes the deuterium abundance relative to hydrogen. For an accretion rate of $\dot{m}=0.1$, \citet{Hosokawa12} demonstrated that deuterium shell burning starts only at protostellar masses of $\sim80$~M$_\odot$, and presumably even higher masses in case of higher accretion rates. A comparison with the Eddington accretion rate in Eq.~(\ref{LEddstar}) thus shows that the luminosity from deuterium shell burning will never become dominant in this mass range, but in fact becomes increasingly less relevant for larger masses. 

It is also clear that deuterium shell burning is not going to occur close to the outer surface. While the characteristic temperature for a given mass shell scales as $T\propto GM/R$, we note that, from Eq.~(\ref{evR}), $GM/R\propto t=M/\dot{M}$ at times $t\gg M/\dot{M}$. In this regime, the central temperature is almost spatially constant, consistent with the results of \citet{Hosokawa12}, but already considerably larger than the deuterium burning temperature of $\sim10^6$~K. Deuterium burning will thus occur when  the first term in Eq.~(\ref{evR}) is still  relevant. From Fig.~\ref{radius}, it is evident that the radius of the shells changes considerably in that regime, while the mass in the shells is almost unchanged. Adopting a scaling relation of $T\propto GM/R$ with almost constant $M$ implies that the radius has to change by 2-3 orders of magnitude for the atmospheric temperature of $5000$~K to increase to a value above $10^6$~K. We therefore expect that the radius of deuterium burning will correspond 
to a fixed fraction of the protostellar radius as long as the protostar maintains its bloated envelope. From a comparison with \citet{Hosokawa12} at a stellar mass of $1000$~M$_\odot$, we obtain the normalization of this relation to be
\begin{equation}
R_{D}=10\, R_\odot \left( \frac{M}{M_\odot} \right)^{1/3}.
\end{equation}
As long as the protostar remains on the Hayashi track, the deuterium burning shell will not be able to move towards the atmosphere, and, therefore, it does not have an impact on the evolution of the protostar.

\begin{figure}[t]
\begin{center}
\includegraphics[scale=0.52]{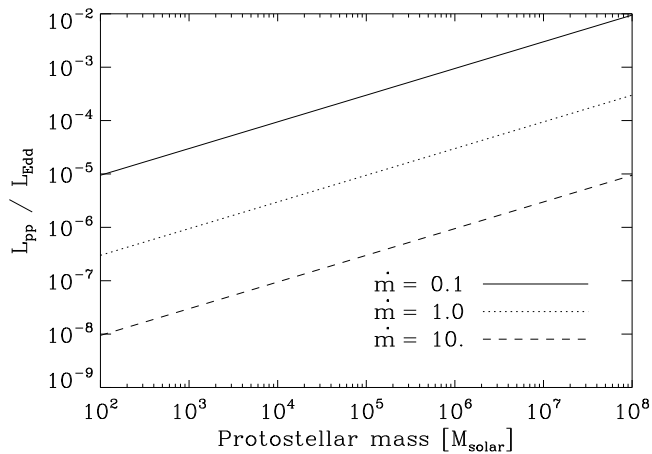}
\caption{The ratio of nuclear luminosity provided via the pp-cycle (\ref{EqLpp}) over the Eddington luminosity of the star (\ref{EqLEdd}) as a function of protostellar mass for different mass accretion rates. Larger accretion times imply that less time was available for contraction, thus reduced nuclear luminosities.  }
\label{Lpp}
\end{center}
\end{figure}

\subsection{The pp-cycle}

We consider first the energy production in the nuclear core via the pp-cycle, as the composition of the star is initially primordial. As shown by \citet{Hosokawa12}, the typical temperature in the core is $\sim10^8$~K, and the nuclear density 
$\sim1$~g~cm$^{-3}$. We adopt here the expression of \citet{Padmanabhan00} for the energy production rate, noting that\begin{equation}
\epsilon_{pp}=\frac{2.4\times 10^4\rho X^2}{T_9^{2/3}}e^{-3.38/T_9^{1/3}}\mathrm{\ erg\ s}^{-1}\mathrm{\ g}^{-1},
\end{equation}
where $T_9$ denotes the temperature in units of $10^9$~K and $X\sim0.75$ the mass fraction of hydrogen. While the pp-cycle considered here has only a moderate temperature dependence, we note that the CNO cycle scales as $T^{20}$ around temperatures of $10^6$~K, and the triple-$\alpha$ process as $T^{40}$ around temperatures of $10^8$~K. For the latter cases, we will thus need to take into account the higher temperatures within the center of the core, as they substantially contribute to the energy production via nuclear burning. Now, with $\rho_{nuc}\sim1$~g~cm$^{-3}$ and $T_9=0.1$, we have $\epsilon_{pp}=43$~erg~g$^{-1}$~s$^{-1}$. The luminosity provided by the pp-cycle is then given as
\begin{equation}
L_{pp}=\epsilon_{pp}M_{nuc}=21.5\, t_{710}^{3/2}\, L_\odot.
\label{EqLpp}
\end{equation}
For comparison, the Eddington luminosity of the protostar is given as\begin{equation}
L_{\rm Edd}=3.8\times10^4 \dot{M} t\,L_\odot=2.7\times10^7\dot{m}\, t_{710} L_\odot .
\label{EqLEdd}
\end{equation}
The ratio of these luminosities is given in Fig.~(\ref{Lpp}), showing that it remains considerably smaller than unity for stellar masses up to $10^8$~M$_\odot$. Equating the two expressions, we find that the luminosity resulting from the pp-burning is relevant only at very late times $t_{pp}=1.1\times10^{15}\ \mathrm{yr}$. 

For  practical accretion rates of $\dot{m}=10^{-3}-10$ and protostellar masses between $10^3-10^8$~M$_\odot$, the contribution from the pp-chain will never be relevant, and can thus be neglected for the overall evolution of the stars. 

We now estimate the amount of helium produced in the core. During one fusion event, an average energy of $4.3\times10^{-5}$~erg is released, implying a helium production rate of
\begin{equation}
\dot{N}_{\mathrm{He,pp}}=\frac{L_{pp}}{4.3\times10^{-5}\mathrm{\ erg}}.%=2.0\times10^{39}\, t_{710}^{3/2}\,\mathrm{s}^{-1}.
\end{equation}
The helium mass in the core thus evolves as\begin{equation}
\dot{M}_{\mathrm{He,pp}}=4m_p\times\dot{N}_{He,pp}.%=2.1\times10^{-10}\,t_{710}^{3/2}\,\frac{M_\odot}{\mathrm{yr}}.
\end{equation}
An integration  yields
\begin{equation}
M_{\mathrm{He,pp}}=6.0\times10^{-8}\,t_{710}^{5/2}\,M_\odot.
\end{equation}
Equating the resulting mass with the total mass in the core, a helium core can be expected after a time of $1.2\times10^{10}$~yr. In summary, when only the pp-cycle is considered, nuclear burning is rather inefficient, and never going to have a significant impact on the evolution of the star. However, we will  show in the next subsection that a transition to the CNO-cycle should be expected, and the implications of nuclear burning will then become more relevant.

\end{document}